\documentclass[sigconf]{acmart}

\usepackage{xspace}
\newcommand{\sysname}{{INDIGO}\xspace}
\newcommand{\BLUE}[1]{\textcolor{black}{#1}}
\newcommand{\codecomment}[1]{\text{\color{gray}// #1}}

\usepackage{textcomp}
\usepackage{pifont}
\usepackage{subcaption}

\usepackage{nicefrac}
\usepackage[utf8]{inputenc}
\usepackage[english]{babel}
\PassOptionsToPackage{bookmarks=false}{hyperref}
\usepackage{multirow, multicol, booktabs, tabulary, tabu, longtable, array, varwidth}
\usepackage{placeins, lipsum}
\usepackage[flushleft]{threeparttable}
\usepackage{tabularx}
\usepackage{amsmath}

\usepackage{textcomp}
\usepackage[shortcuts,acronym]{glossaries}
\usepackage{soul, footnote, xargs}

\usepackage{amsmath}
\usepackage[utf8]{inputenc}
\usepackage{enumerate}

\usepackage[inline]{enumitem}
\setlist{noitemsep,nolistsep,leftmargin=*}

\usepackage{amsmath, amsfonts, amssymb, amsthm, nicefrac}
\theoremstyle{definition}

\usepackage{listings}

\usepackage[scaled=0.82]{beramono}

\usepackage[]{cleveref} 
\crefname{appsec}{Appendix}{Appendices}
\crefformat{section}{\S#2#1#3}
\crefformat{subsection}{\S#2#1#3}
\crefformat{subsubsection}{\S#2#1#3}
\crefformat{equation}{(#2#1#3)}
\crefrangeformat{equation}{(#3#1#4--#5#2#6)}
\crefmultiformat{equation}{(#2#1#3)}{ and~(#2#1#3)}{, (#2#1#3)}{ and~(#2#1#3)}
\crefformat{figure}{Fig.~#2#1#3}
\crefrangeformat{figure}{Figs. #3#1#4--#5#2#6}
\crefmultiformat{figure}{Figs.~#2#1#3}{ and~#2#1#3}{, #2#1#3}{ and~#2#1#3}
\crefname{algocf}{Alg.}{Algs.}
\Crefname{algocf}{Algorithm}{Algorithms}
\crefformat{table}{Table~#2#1#3}
\crefrangeformat{table}{Tables~#3#1#4--#5#2#6}
\crefmultiformat{table}{Tables~#2#1#3}{ and~#2#1#3}{, #2#1#3}{ and~#2#1#3}

\usepackage{textcomp}
\usepackage[shortcuts,acronym]{glossaries}
\usepackage{soul, footnote, xargs}
\usepackage[colorinlistoftodos,prependcaption,textsize=tiny]{todonotes}
\usepackage{balance}

\usepackage{graphicx}
\usepackage{tikz, epstopdf, stfloats, bbding, capt-of}

\usepackage{algorithm}
\usepackage{algpseudocode}

\usepackage{microtype}
\usepackage{xspace}
\usepackage[T1]{fontenc}
\usepackage[compact]{titlesec}
\titlespacing*{\section}{0pt}{6pt plus 4pt minus 2pt}{2pt plus 2pt minus 2pt}
\titlespacing*{\subsection}{0pt}{4pt plus 2pt minus 1pt}{2pt plus 1pt minus 1pt}
\titlespacing*{\subsubsection}{0pt}{4pt plus 2pt minus 1pt}{2pt plus 1pt minus 1pt}

\AtBeginDocument{%
  }

\setcopyright{none} 
\copyrightyear{2025}
\acmYear{2025}
\acmDOI{XXXXXXX.XXXXXXX}

\acmConference['ICS 2025']{The 39th ACM International Conference on
  Supercomputing}{June 9--11, 2025}{Salt Lake City, USA}
\acmISBN{000-0-0000-0000-0/00/00}

\acmSubmissionID{\#166}

\begin{document}

\title[Page Migration for HMD Systems]{\Large \bf \sysname: Page Migration for Hardware Memory Disaggregation Across a Network}

\author{Archit Patke}
\affiliation{%
  \institution{University of Illinois at Urbana-Champaign}
  \city{Urbana}
  \state{Illinois}
  \country{USA}
}
\author{Christian Pinto}
\affiliation{%
  \institution{IBM Research}
  \city{Dublin}
  \country{Ireland}
}
\author{Saurabh Jha}
\affiliation{%
  \institution{IBM Research}
  \city{Yorktown Heights}
  \state{New York}
  \country{USA}
}
\author{Haoran Qiu}
\affiliation{%
  \institution{University of Illinois at Urbana-Champaign}
  \city{Urbana}
  \state{Illinois}
  \country{USA}
}
\author{Zbigniew Kalbarczyk}
\affiliation{%
  \institution{University of Illinois at Urbana-Champaign}
  \city{Urbana}
  \state{Illinois}
  \country{USA}
}
\author{Ravishankar Iyer}
\affiliation{%
  \institution{University of Illinois at Urbana-Champaign}
  \city{Urbana}
  \state{Illinois}
  \country{USA}
}

\begin{abstract}
Hardware memory disaggregation (HMD) is an emerging technology that enables access to remote memory, thereby creating expansive memory pools and reducing memory underutilization in datacenters.
However, a significant challenge arises when accessing remote memory over a network: increased contention that can lead to severe application performance degradation.
To reduce the performance penalty of using remote memory, the operating system uses \emph{page migration} to promote frequently accessed pages closer to the processor.
However, previously proposed page migration mechanisms do not achieve the best performance in HMD systems because of obliviousness to variable page transfer costs that occur due to network contention.
To address these limitations, we present \sysname: a network-aware page migration framework that uses novel page telemetry and a learning-based approach for network adaptation.
We implemented \sysname in the Linux kernel and evaluated it with common cloud and HPC applications on a real disaggregated memory system prototype.
Our evaluation shows that \sysname offers up to 50--70\% improvement in application performance compared to other state-of-the-art page migration policies and reduces network traffic up to 2\texttimes.

\end{abstract}

\maketitle

\section{Introduction}
\label{s:introduction}

\begin{figure}[t!]
    \includegraphics[width=3.3in]{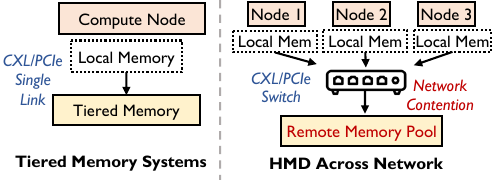}
\vspace{1ex}
\caption{Compared to tiered memory systems, multiple nodes contend for the same remote memory pool leading to network contention in HMD systems.}
\label{fig:intro_degradation}
\end{figure}

\noindent \textbf{Background and Motivation.}
Hardware memory disaggregation (HMD) is an emerging technology that enables cache-line granularity access to remote memory with protocols such as CXL or openCAPI~\cite{CXL,openCAPI,CCIX}, thereby creating expansive memory pools and reducing memory underutilization in datacenters.
To cost effectively support the growing requirements of memory-intensive applications~\cite{stonebraker2013voltdb,vaswani2017attention,murphy2010introducing}, memory expansion has gone beyond a single compute node to multiple nodes and a remote memory pool that are connected together with a switched interconnect~\cite{pinto2020thymesisflow,xconn-switch,h3-platform}.
As with any switched network, contention from multiple hosts results in slow down of remote memory access as shown in Figure~\ref{fig:intro_degradation}.
Specifically, we find that multi-node contention in HMD systems results in 3--6\texttimes{} higher performance degradation compared to traditional tiered memory systems~\cite{gouk2023memory,gouk2022direct,patke2022evaluating,lim2009disaggregated,syrivelis2017software}.
Our discussions with a leading cloud and HPC provider indicate that network contention is a significant bottleneck in the deployment of CXL memory.

\begin{figure}[ht]
    \centering
        \hspace{-3ex}
    \begin{minipage}{.1\textwidth}
      \includegraphics[width=1.7in]{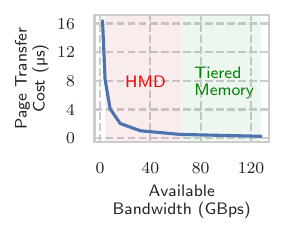}
    \end{minipage}
    \hspace{18ex}
    \begin{minipage}{.2\textwidth}
      \includegraphics[width=1.7in]{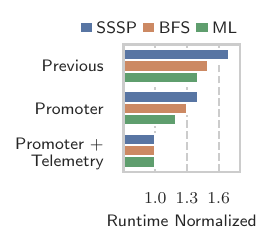}
    \end{minipage}
    \caption{Previously proposed page migration mechanisms do not consider the variable page transfer costs in HMD systems. (Left) Page transfer cost for 4 KB pages in tiered memory and HMD systems under varying network contention. (Right) \sysname considers the impact of variable page transfer costs with Page Promoter and Page Telemetry resulting in decreased application runtime. Each application was run with local memory allocation = 10\% of the working set size and compared with the best runtime between TPP~\cite{maruf2022tpp}, MEMTIS~\cite{memtis_sosp} and Nimble~\cite{yan2019nimble}}
    \label{fig:page_transfer_cost}
\end{figure}

\emph{Page migration}~\cite{linux_migration} can be reused to address network contention in HMD systems by transferring frequently accessed pages from remote to local memory.
However, the associated \emph{page transfer cost}, i.e. the time required to copy pages between remote and local memory, is higher in HMD systems and increases during network contention as shown in Figure~\ref{fig:page_transfer_cost} (Left).
Previously proposed page migration mechanisms such as ~\cite{linux_migration,kim2021exploring,verghese1996operating,agarwal2017thermostat,ren2023hm,kannan2017heteroos,raybuck2021hemem,li2023hopp,yan2019nimble,maruf2022tpp,memtis_sosp} that are designed for legacy NUMA systems, or disaggregated systems with point-to-point CPU-memory connections, do not consider such variable page transfer costs resulting in sub-optimal scenarios as described below.

First, previously proposed mechanisms are not effective at transferring pages during network contention.
Page migration is only beneficial when the improvement in access locality outweighs the page transfer cost.
During network contention, the cost of transferring pages increases because of reduced available bandwidth, which results in excessive page transfers with the current mechanisms.
    
Second, previously proposed mechanisms fail to transfer pages optimally when application access patterns are dynamic.
Page migration is based on access rates estimated using moving averages~\cite{raybuck2021hemem,memtis_sosp} or bit vectors~\cite{kwon2016coordinated}.
These estimates become highly inaccurate with dynamic patterns.
For example, Graph500 applications like BFS and SSSP have estimation errors greater than 1000\%, resulting in the transfer of infrequently accessed pages.
In HMD systems, this issue is exacerbated due to the high cost of page transfers.

\noindent \textbf{\sysname Design.}
To resolve the limitations of previous page migration mechanisms in the context of HMD systems, we introduce \sysname.
\sysname consists of two major components: \emph{Page Telemetry} and \emph{Page Promoter}.

\emph{Page Telemetry} improves the accuracy of page access estimation by differentiating between stationary and dynamic application access patterns.
Differentiation between the two patterns is enabled by calculating the size of each \textit{interval} (i.e., ``burst duration'') in which the access pattern does not vary significantly.
Smaller burst durations imply that the application memory access pattern is dynamic (and vice versa for larger ones).

\emph{Page Promoter} adopts a network-aware page migration policy that considers the trade-off between the cost of page transfer and memory access locality benefit (we develop an analytical model to consider this trade-off in Section~\ref{ss:network_cost_model}) within the constraint of limited local memory allocation.
The input to the Page Promoter includes measurements from Page Telemetry, application memory allocation, and network telemetry.
The Page Promoter migrates pages if the access rates and burst duration exceeds a threshold value.
This threshold value is decided at the application start by a contextual multi-armed bandits agent that is trained using historical application runtime data.
We also provide an alternate heuristic-based network adaptive policy if no application training data is available\footnote{The heuristic-based approach is only network adaptive but not adaptive to the dynamicity of application access patterns}.

\sysname is implemented between Linux kernel and the userspace.
Page Telemetry and Page Promoter are both implemented in the Linux kernel, while the contextual multi-armed bandits control is implemented in the userspace.
Both components contribute to \sysname's improved performance as shown in Figure~\ref{fig:page_transfer_cost} (Right).

\noindent \textbf{Results.}
We evaluated \sysname with common cloud applications, including in-memory databases~\cite{stonebraker2013voltdb}, in-memory key-value stores~\cite{carlson2013redis}, ML inference~\cite{vaswani2017attention}, graph analytics kernels~\cite{murphy2010introducing}, and scientific computational code~\cite{gahvari2011modeling} on a real HMD prototype system: ThymesisFlow~\cite{thymesisflow}.
We compared INDIGO with state-of-the-art page migration techniques and recent research work available in the literature, such as TPP~\cite{maruf2022tpp}, Nimble~\cite{yan2019nimble}, and MEMTIS~\cite{memtis_sosp}.
We summarize key results below:
\begin{enumerate}[label=\arabic*.,wide=0pt]
    \item \emph{Improved Application Performance.} \sysname improves application performance by up to 50--70\% compared to other state-of-the-art page migration policies across low to high local-to-remote memory allocation ratios. The network adaptive policy also offers an improvement of up to 20--50\% in the presence of network contention.
    \item \emph{Unlocking Memory Offload.}
    For applications such as ML inference, \sysname can match full local memory execution performance with limited local memory allocation, thereby enabling 80--90\% offload of the application working set\footnote{The ``working set'' of an application is the collection of memory pages that are currently owned by the application.} from local to remote memory.
    \item  \emph{Reduced Network Traffic.} \sysname reduces network traffic by up to 3\texttimes{} compared to other policies, such as MEMTIS, TPP, and Nimble, while improving the overall application performance. 
    The decreased network traffic leads to reduced contention and improved colocation for multiple applications that share the network. 
    \item \emph{Minimal Training Costs.}
    Our evaluation shows that \sysname, when exclusively trained on a particular application, will perform equally well on other, similar applications without any retraining.
    For example, there is a $<$5\% drop in application performance when one trains \sysname on a breadth-first search (BFS) application but tests the trained model on a single source shortest path (SSSP) application (with no retraining/adaptation).
\end{enumerate}
\noindent \textbf{Putting \sysname in perspective.}
The problem of deciding the location of memory pages is also present in other OS-based approaches to memory disaggregation that use RDMA along with a swap cache such as InfiniSwap~\cite{gu2017efficient}, Leap~\cite{al2020effectively},
Fastswap~\cite{amaro2020can},
and Hermit~\cite{qiao2023hermit}.
However, these RDMA-based systems are fundamentally different from HMD systems, as they emulate disaggregated memory via the kernel; rather than extending the address space.
Consequently, \emph{they do not have the option of keeping the accessed page in remote memory};  it has to be brought to local memory.
That changes the decision problem, and the analytical model (later presented in Section~\ref{ss:network_cost_model}) no longer holds. 
The page-to-transfer decision becomes a prefetching problem as opposed to a migration problem.

\section{Background}
\label{s:background}
We discuss the background of HMD systems and page migration, and highlight the drawbacks of state-of-the-art page migration mechanisms when applied to HMD systems.


\subsection{Hardware Memory Disaggregation}
\label{ss:background_hmd}
\sysname targets hardware memory disaggregation (HMD), which employs cache lines as a basic block for memory access.
Similar to tiered memory, the remote memory pool is accessed only in response to a last-level CPU cache miss, without any intervention of the operating system or any other software runtime.
Cache lines are transmitted via a high-performance cache coherence protocol such as OpenCAPI~\cite{openCAPI} or CXL~\cite{CXL}.
Additionally, HMD systems have multiple nodes accessing the same remote memory pool as shown in Figure~\ref{fig:intro_degradation}.
Due to cache line-level access, HMD exhibits the lowest remote memory access latency as it eliminates the software overhead present in other OS-based~\cite{shan2018legoos,gu2017efficient,aguilera2018remote,amaro2020can,gao2016network,koussih1999dodo,liang2005swapping,feeley1995implementing,lee2022hydra} or custom runtime-based approaches~\cite{wang2020semeru,ruan2020aifm,zhou2022carbink} for disaggregation.
Regardless of its implementation, accessing of remote memory is generally slower than accessing of local memory due to the higher access latency and lower bandwidth availability that result from the memory network, and performance degradation is expected for most applications.
Most HMD solutions~\cite{pinto2020thymesisflow,calciu2021rethinking} reuse the existing OS memory features created specifically for the NUMA architecture and materialize disaggregated memory as a CPU-less remote NUMA node.
We adopted the same design choice for \sysname.

\subsection{Network Contention}
Memory contention is a well-known phenomenon in computing systems, where multiple application streams or processor cores access memory devices across shared links, sharing the fixed amount of bandwidth available to transfer data to/from memory.
Memory contention would be even worse in HMD systems based on a switched network, where a number of nodes access the remote memory pool via one or more switches~\cite{cxl-memory-expansion,xconn-switch}.
For example, the commercial Falcon C5022 from H3 Platform, based on a CXL 2.0 switch~\cite{h3-platform}, enables sharing a single memory device over a PCIe 5.0 x8 link (64GBps) with up-to 8 compute nodes, resulting in an effective bandwidth of 8 GBps per node when they access memory locations in the same device simultaneously.

\subsection{Page Migration}
\label{ss:background_migration}

\begin{table*}[]
\resizebox{1\linewidth}{!}{
\begin{tabular}{@{}cccccc@{}}
\toprule
\bf System      & \bf Access Estimation & \bf Migration Policy & \bf Network Adaptation & \bf E2E Performance Aware & \bf Page Demotion \\ \midrule
autoNUMA~\cite{intel-autonuma}   & \ding{55}                & Static Threshold                   & \ding{55}                 & \ding{55}           & Recency                   \\
autoTiering~\cite{kim2021exploring} & \ding{55}                & Promotion-based      & \ding{55}                 & \ding{55}           & Recency                   \\
TPP~\cite{maruf2022tpp}         & \ding{55}                & Static Threshold                   & \ding{55}                 & \ding{55}           & Recency                   \\
Nimble~\cite{yan2019nimble}      & \ding{55}                & Static Threshold                   & \ding{55}                 & \ding{55}           & Recency                   \\
Multi-Clock~\cite{maruf2022multi} & \ding{55}                & Static Threshold                   & \ding{55}                 & \ding{55}           & Recency + Frequency                   \\
HeMem~\cite{raybuck2021hemem}       & Moving Average    & Static Threshold                   & \ding{55}                 & \ding{55}           & Frequency                   \\
MEMTIS~\cite{memtis_sosp}      & Moving Average    & Histogram-Based      & \ding{55}                 & \ding{55}           & Recency + Frequency                   \\ \midrule
\textbf{INDIGO (w/o training)}     & \ding{55}    &  Network-Based      & \ding{51}                 & \ding{55}           & Recency + Frequency \\
\textbf{INDIGO (with training)}     & \multicolumn{2}{c}{Contextual Bandits}       & \ding{51}                & \ding{51}           & Recency + Frequency                  \\ \bottomrule
\end{tabular}
}
\vspace{1ex}
\captionof{table}{Comparison of \sysname with other existing page migration mechanisms in disaggregated and tiered memory systems.}
\label{tab:comparison_page_migration}
\end{table*}

A memory page is the logical abstraction of memory at the operating system level.
The physical location of each page can be dynamic and can vary between nodes in a NUMA system such that frequently accessed memory is closer to the CPU in which the computation is taking place (local).
In this approach, unlike disk-based swapping, a page in remote memory is accessible to the CPU at cache line granularity, albeit with higher latency.
To decide the location of each page dynamically, the operating system implements a \emph{page migration} policy that consists of
\begin{enumerate*}[label=(\alph*)]
\item page telemetry, which measures page access patterns; 
\item page promotion, which transfers pages from remote to local memory; and
\item page demotion, which transfers pages from local to remote memory.
\end{enumerate*}
As page access rate may vary over time, page promotion and demotion are always ``on'' and work together to dynamically decide on whether to place pages in the remote memory pool or on local nodes.

\subsection{Network Cost Model for Page Migration}
\label{ss:network_cost_model}
We introduce a network cost model for page migration in HMD systems and leverage it to justify our design choices for \sysname throughout the rest of this paper.

The model considers a migration (i.e., swap) between two pages \textbf{p} (promoted page) and \textbf{d} (demoted page) initially located in remote and local memory respectively.
Page migration benefits memory access locality (bringing it closer to compute) for page \textbf{p} while harming access locality for page \textbf{d} and costing time for the page transfer on the network.
The advantage of page migration is only realized when the benefit is more than the cost as represented in Equation~\cref{eq:axiom_page_migration}.
\begin{equation}
\underbrace{\int_{0}^{\Delta T} (\text{access}_p - \text{access}_d) \: dt}_{\text{Access Locality Benefit}} \geq \underbrace{\frac{\text{page size}}{\text{bandwidth} \times \Delta \: \text{latency}}}_{\text{Page Transfer Threshold}} + k
\label{eq:axiom_page_migration}
\end{equation}
$\Delta T$ is the interval between the current and next swaps of the pages and is determined by the page migration policy.
$access_p$ and $access_d$ are the access rates for the pages $p$ and $d$ during the interval $\Delta T$.
The network cost of page transfer is represented by $\frac{\text{page size}}{\text{bandwidth} \times \Delta \: \text{latency}}$, where $\Delta \: \text{latency}$ is the difference between remote and local memory latency.
$k$ is the constant associated with the bookkeeping overhead of page migration that includes handling interrupts, remapping of page table entries, flushing of the translation lookaside buffer (TLB), and so on.



\subsection{Limitations of Existing Page Migration Mechanisms}
\label{ss:page_migration_characterization}

Previous work on page migration has adopted different approaches for page telemetry, promotion, and demotion.
We summarize these approaches in~\cref{tab:comparison_page_migration}.

\noindent \textbf{Page Telemetry.} The ideal migration decision is based on an estimate of the difference between the access patterns for the promoted and demoted pages ($\int_{0}^{\Delta T} \text{access}_p dt$ and $\int_{0}^{\Delta T} \text{access}_d dt$ in Equation~\cref{eq:axiom_page_migration}) and whether it exceeds the cost of transferring the page.
Other than MEMTIS, no mechanism considers the access rate of the demoted page ($\int_{0}^{\Delta T} \text{access}_d dt$) \BLUE{during decision-making for page promotion.}
Instead, they rely solely on an estimate of the access rate of the promoted page.
However, such an approach can lead to the promotion of pages even when the access rate of the demoted page is high.
This scenario is suboptimal as the benefit of page migration (shown on the left-hand side of Equation~\cref{eq:axiom_page_migration}) is low or even negative, thereby causing performance loss.
Even for estimating the access rate of the promoted page ($\int_{0}^{\Delta T} \text{access}_p dt$), mechanisms like autoNUMA, autoTiering, TPP, Nimble, and Multi-Clock use the last access measurement without normalization or prediction.
Others, such as HeMem and MEMTIS, use variations of exponentially weighted moving averages (EWMA) to smooth out previous measurements.
However, EWMA may not be accurate for dynamic workloads. (We characterize this problem in Insight 2, below.)
In addition, it is challenging to determine the appropriate decision context (i.e., $\Delta \: T$) to accurately estimate the relevant time window for the estimated access patterns.

\noindent \textbf{Page Migration.} Fundamentally, most previous mechanisms are suboptimal in HMD systems because they do not consider the page transfer cost term in Equation~\cref{eq:axiom_page_migration}. (We characterize this problem in Insight 1, below.)
Mechanisms such as autoNUMA, TPP, Nimble, and HeMem have static thresholds for promotion (such as promoting if $\int_{0}^{\Delta T} \text{access}_p dt > \text{fixed constant}$).
In particular, autoTiering and MEMTIS adapt thresholds based on different criteria.
autoTiering migrates the most recently accessed pages to meet a target promotion rate.
MEMTIS also considers the access rate of the demoted page and migrates pages such that the local memory allocation is always at capacity.
Neither of those mechanisms can easily be modified to support network cost adaptation.

\begin{figure*}
\begin{minipage} {0.32\textwidth}
\includegraphics{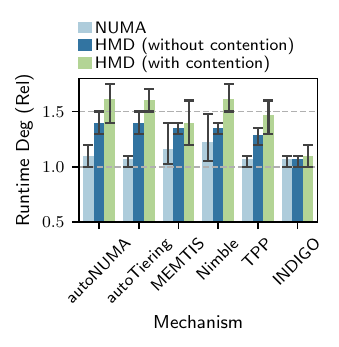}
\caption{Runtime degradation for different page migration mechanisms.}
\label{fig:insight_runtime_degradation}
\end{minipage}
\hfill
\begin{minipage} {0.32\textwidth}
\includegraphics{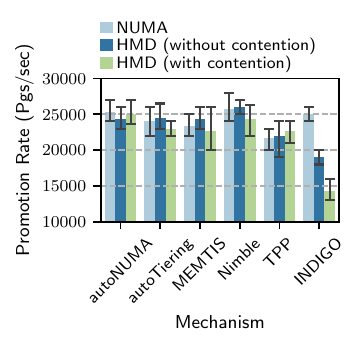}
\caption{Page promotion rates with different page migration mechanisms.}
\label{fig:insight_promotion_rate}
\end{minipage}
\hfill
\begin{minipage} {0.32\textwidth}
\includegraphics{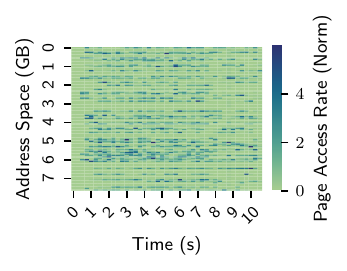}
\caption{Dynamically shifting application access patterns for a BFS application.}
\label{fig:ewma_shortcomings}
\end{minipage}
\hfill

\end{figure*}


\subsection{Core Insights}
\label{ss:core_insights}

The above mentioned background knowledge and limitations of prior approaches led to three core insights, that we use to motivate the design of \sysname.

\noindent
\textit{\textbf{Insight 1: Existing page migration mechanisms fail to adapt based on changing network conditions.}}
Previous work on page migration did not take into account the network cost of transferring pages: $\frac{\text{page size}}{\text{bandwidth} \times \Delta \: \text{latency}}$ (as shown in Equation~\cref{eq:axiom_page_migration}).
This network cost term can be up to 16$\times$ higher in HMD systems than in conventional NUMA architecture because of the smaller network bandwidth (especially during multi-tenant contention).
Consequently, mechanisms that do not consider network cost perform well for conventional NUMA but make suboptimal decisions when ported to HMD systems, e.g., transferring excessive amounts of pages during periods of network contention.

To characterize that behavior, we ran the Graph 500 BFS application with 5 different page migration mechanisms in three scenarios:
\begin{enumerate*}[label=(\alph*)]
    \item a high-bandwidth local NUMA node (NUMA) equivalent to tiered memory,
    \item \BLUE{a remote memory pool without additional network contention (HMD without contention),} and 
    \item a remote memory pool under network contention via a memory \BLUE{copy} benchmark (HMD with contention)~\footnote{STREAM~\cite{bergstrom2011measuring} COPY kernel was used to reduce available bandwidth to 8 GBps to create multi-node network contention.}.
\end{enumerate*}
Figure~\ref{fig:insight_runtime_degradation} shows the relative runtime degradation (normalized by the best run of the best-performing mechanism. Coloured bars show the median value, while error bars show the range min-max across all runs) of the BFS application under the three scenarios.
While most of the page migration mechanisms performed well on the local NUMA node, \BLUE{their performance suffered in the HMD system: both with and without network contention.}
The performance improvement offered by \sysname can be explained using Equation~\cref{eq:axiom_page_migration}.
Network access causes the cost of migrating pages ($\frac{\text{page size}}{\text{bandwidth} \times \Delta \: \text{latency}}$) to increase significantly because of the lower available bandwidth (100$\times$ lower than for NUMA in our prototype).
Consequently, the difference between the access rates of the promoted page and the demoted page ($\text{access}_p - \text{access}_d$) needs to be higher for optimal transfer.
Therefore, fewer pages would have access rates that are high enough to justify migration and the overall promotion rate should be lower.
Characterization of the promotion rates as shown in Figure~\ref{fig:insight_promotion_rate} confirms this observation.
\sysname is more selective in page promotion, as the promotion rate is about 25--50\% lower than that of the other mechanisms.
\vspace{0.5ex}

\noindent\fbox{%
    \parbox{\columnwidth}{%
        Page migration mechanisms must adapt their page promotion selectivity based on the network conditions in HMD systems.
    }%
}
\vspace{0.5ex}



\noindent
\textit{\textbf{Insight 2: Relying solely on moving averages is insufficient for page access telemetry.}}
Page migration mechanisms such as HeMem and MEMTIS use exponentially weighted moving averages (EWMA) to estimate future memory access patterns based on historical access records and use them to make the migration decision.
EWMA can be very effective at dealing with \textit{stationary} memory access patterns, as minor measurement fluctuations are corrected.
However, we find that it can be completely ineffective when dealing with dynamically shifting application access patterns.
For instance, consider the access patterns of a segment of the Graph 500 BFS application as shown in Figure~\ref{fig:ewma_shortcomings}.
As the page access patterns keep shifting dynamically, EWMA is unable to estimate which pages will be frequently accessed causing them to remain in the remote memory pool.
For example, EWMA with $\alpha = 0.5$ (used in HeMem) and $\alpha = 0.9$ (used in MEMTIS) leads to p75 error = 170\% and p99 error = 60000\% 
when estimating the access rate in the next measurement window.
\vspace{0.5ex}

\noindent\fbox{%
    \parbox{\columnwidth}{%
        There is a need to redesign page telemetry so that it can work well for both \textit{stationary} and \textit{dynamic} application access patterns.
    }%
}
\vspace{0.5ex}

\noindent
\textit{\textbf{Insight 3: Accurate estimation of page access patterns alone is not sufficient for the page migration decision problem.}}
\BLUE{Equation~\cref{eq:axiom_page_migration} shows that optimal decision-making depends on estimating $\int_{0}^{\Delta T} (\text{access}_p - \text{access}_d) \: dt$; that involves the decision context ($\Delta T$) in addition to the page access rates.
The optimization problem of selecting the correct $\Delta T$ is difficult to solve optimally online.}
A possible solution is to use algorithmic optimization techniques (such as bipartite matching) paired with a supervised learning model that estimates page access patterns.
However, such an optimization requires computation of $\sim O(\text{pages}^3)$~\cite{kuhn1955}, which is prohibitively expensive for large memory allocations (i.e., minutes $\gg$ sub-millisec).

\noindent\fbox{%
    \parbox{\columnwidth}{%
        Classical algorithmic optimization leads to prohibitively high computational costs for making online page migration decisions.
    }%
}
\section{\sysname Design Overview}
\label{s:design}

\noindent
We derived the design of \sysname from first principles driven by the insights presented in Section~\ref{ss:core_insights} and Equation~\cref{eq:axiom_page_migration}.
The page migration decision problem relies on estimates of page access rates ($\text{access}_p$ and $\text{access}_d$) to migrate pages for the decision context ($\Delta T$).
Our design focuses on increasing the estimation accuracy of page access rates via improvements in page telemetry and by optimally solving the decision problem using learning-based techniques.

\noindent \textbf{Design Choice 1: Augmenting page telemetry with contextual (time-based) features for accurate page access estimation.}
The page migration decision is based on estimating the access rates for both the promoted and demoted pages.
Previous mechanisms, such as HeMem and MEMTIS, used exponentially weighted moving averages (EWMA) to estimate these access rates.
However, as we show in Insight 2 in Section~\ref{ss:core_insights}, moving averages fail to adapt effectively, especially when application access patterns vary dynamically.
On the other hand, moving averages work well with static access patterns when small deviations in access rates can be ignored.
Our goal is to design a mechanism that can generalize for both access patterns.
To do so, we introduce a new telemetry feature that captures variability in historic access patterns via a ``burst duration.''
The burst duration is the time interval for which the current access pattern remained valid in the past.
A smaller burst duration indicates dynamic application access patterns (and vice versa for a larger burst duration).
The burst duration adds a time-based dimension to the access measurements necessary for accurate estimation.
In Section~\ref{s:telemetry}, we show that burst duration is easy to compute within the kernel space and adds minimal overhead.

\noindent \textbf{Design Choice 2: Utilizing a feedback loop to solve the page migration optimization problem.}
In addition to accurate estimations of future page access rates, the page migration decision problem also requires selection of the decision context (i.e., $\Delta T$).
As discussed in Insight 3 in Section~\ref{ss:core_insights}, selecting the appropriate $\Delta T$ is difficult via supervised learning because of the high computational overhead ($O(\text{pages}^3$)). 
As solving the vanilla decision problem is challenging, we simplify it by using additional feedback from application performance.
This feedback can be combined with page telemetry to create an end-to-end policy-training loop for the decision optimization problem.
The specific optimizer we chose is contextual multi-armed bandits, as it requires less parameter tuning (compared to out-of-the-box reinforcement learning).
We describe details of the optimization loop in Section~\ref {s:promotion_demotion_demotion}.

\noindent \textbf{Design Choice 3: Enabling dynamic policy adaptation based on application access patterns, network contention, and memory allocation.}
As described in Design Choice 2, we use an end-to-end feedback loop to model and solve the page migration decision problem.
We revisit Equation~\cref {eq:axiom_page_migration} to decide on the necessary input for this feedback loop.
The term $\text{access}_p$ is typically measured by page table poisoning and interception of the corresponding page fault.
This approach is commonly used in other mechanisms, such as autoNUMA, autoTiering, and TPP, and we use it too.
Along with the burst duration, it can be effective for estimating both static and dynamic access patterns (for $\text{access}_p$). 
The other term $\text{access}_d$ is more challenging to estimate as it is not in the critical path of page migration, and page table-based estimates will lag the true access rates.
To resolve that problem, we observed that $\text{access}_d$ is dependent on the (local) memory allocation for a given application.
Pages are ranked by their access rate to decide which ones are eligible for promotion or demotion, and in general those with the lowest access rates would be demoted.
Consequently, the larger the local memory available, the larger is the set of pages that are eligible for promotion and the lower is the access rate of the demoted page. 
Finally, we must take into account the network characteristics that we measure using network hardware performance counters.
By taking into account those features, driven by the formulation presented in Equation~\cref{eq:axiom_page_migration}, \sysname is able to dynamically adapt the page migration policy.

\begin{figure}[!t]
    \includegraphics{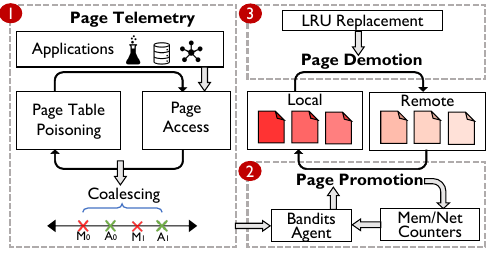}
\caption{\sysname Workflow}
\label{fig:Indigo}
\end{figure}

\noindent \textbf{\sysname Workflow.}
Figure~\ref{fig:Indigo} presents a high-level workflow of \sysname that uses the design choices discussed here.
First, \sysname Page Telemetry measures page access rates and coalesces similar measurements to estimate burst durations.
The burst duration aids in differentiating between static and dynamic access patterns.
Next, these measurements are used along with other system metrics, such as memory allocation and network telemetry (collected from memory/network counters, as shown in the figure), to determine the pages that should be migrated.
The page migration decision is then taken by a contextual bandit agent trained in a feedback loop using application performance metrics (for e.g. application completion time).
If the collection of training data is not feasible, \sysname also offers a network adaptive policy that adjusts promotion rates based on varying network conditions.  
Finally, infrequently accessed pages are demoted back to the remote memory pool using the regular Linux LRU replacement mechanism.
\section{Page Telemetry}
\label{s:telemetry}
In \sysname, the main goal of Page Telemetry is to enable the estimation of page access patterns to be used in the feedback loop.
The Page Telemetry is primarily based on page table poisoning similar to  mechanisms such as autoNUMA, TPP, and autoTiering.
We use additional enhancements to address the shortcomings of other mechanisms discussed in Section~\ref{ss:page_migration_characterization}:
\begin{enumerate*}[label=(\alph*)]
    \item better estimation of access rate, and
    \item a contextual (time-based) feature for page accesses with \textit{burst duration}.
\end{enumerate*}

\subsection{Page Table Poisoning}
\label{ss:page_table}

At an interval of one second, \sysname periodically marks all pages of an active process to be inaccessible and records the \emph{marking time} ($M_i$).
If a page has already been marked in the current interval (but not accessed), it is not marked again in the next interval.
When a process is accessing memory for the first time after the marking, a page fault (known as a \emph{NUMA hinting fault}~\cite{intel-autonuma}) is generated that can be intercepted to record the corresponding \emph{access time} ($A_i$).
After each access, pages are unmarked to prevent additional faults in that interval.
Using this method, one can generate a timeline of marking and access times for each page. We use that timeline to estimate the page access rate.

\subsection{Point Access Rate Estimator}
\label{ss:access_access rate}

To estimate the page access rate, we use a combination of a pair of marking and access times.
One can see intuitively that if a page is accessed immediately after it is marked (i.e., the difference between the access and marking times is small), its access rate will likely be high.
Conversely, if a page is marked but not accessed for a long duration of time (i.e., the difference between the access and marking times is large), its access rate would be low.

\begin{equation}
    F_i = 1/(A_i-M_i)
\label{eq:access rate_estimation}
\end{equation}

We use Equation~\cref{eq:access rate_estimation} to estimate $F_i$, the access rate for the $i$-th marking and access time pair.
Note that $A_i$ will always be greater than $M_i$, as access occurs after marking.

\subsection{Coalescing Access Measurements}
\label{ss:burst_duration}

\begin{algorithm}[!t]
\caption{Coalescing Access Measurements to Estimate Burst Duration (i.e., the Size of the Cluster).}
\label{algo:clustering-access-access rate}
\begin{algorithmic}[1]
\Require Marking and Access Times $\langle M,A \rangle$
\Procedure{ClusterAccessRate}{$\langle M,A \rangle$}
    \State Clusters $C \gets \varnothing$
    \State Access Rate $F \gets \varnothing$
    \For {each $\langle M_i,A_i \rangle$ in [$\langle M,A \rangle$]}
        \State $F_i = 1/(A_i-M_i)$
        \State $sz$ = size($C$)
        \If {$i>0 \And |F_i - F_{i-1}| < \delta1 \And M_i - M_{i-1} \leq \delta_2$}
            \State $C_{sz-1}$.add(i) \codecomment{Add access rate to curr cluster}
        \Else
            \State $C$.append({}) \codecomment{Create a new cluster}\phantom{Add access rate to current cluster}
            \State $C_{sz}$.add(i) \codecomment{Add access rate to new cluster}
        \EndIf
	\EndFor
	\State \Return Clusters $C$, Access Rate $F$
\EndProcedure
\end{algorithmic}
\end{algorithm}

Based on Insight 2 in Section~\ref{ss:core_insights}, a single access rate measurement based on a moving average is insufficient to capture the temporal behavior of page access patterns.
To capture the access patterns across time, we also group measurements with similar access frequencies together (i.e., a \textit{burst}) to estimate the duration of access (i.e., the \textit{burst duration}).
We can use an online segmentation algorithm to perform this grouping.
Algorithm~\ref{algo:clustering-access-access rate} is one such segmentation algorithm that estimates the burst duration for page access; \BLUE{$\delta_1$ denotes the closeness of two access rate measurements and $\delta_2$ is simply the page marking interval.
The values for $\delta_1$ and $\delta_2$ are selected with Bayesian optimization.}
For each page, we maintain an additional variable in the page struct to denote the current size of the cluster/burst.
If the next access rate measurement is close in magnitude to the previous measurement and the measurement interval (i.e., the duration between two adjacent marking times) is small, then the next measurement is added to the current cluster.
Otherwise, we create a new cluster and start collecting the next measurement.
The size of each cluster effectively represents the burst duration based on the first and last measurements added in the cluster.
Overall, the clustering output from the algorithm would be a tuple of the burst duration $C$ and the estimated access rate $F$ for every page across all measurements.
The hyper-parameters in page telemetry such as page marking time and closeness thresholds are determined empirically.

\BLUE{As an alternative to page marking, PEBS counters can also be utilized to determine page access rates.
However, \sysname does not use PEBS because:
\begin{enumerate*}[label=(\alph*)]
    \item PEBS-based profiling leads to increased overheads for workloads with high memory bandwidth requirements~\cite{maruf2022tpp}, and
    \item limited number of perf counters are supported in CPUs that may be required for other system monitoring tasks.
\end{enumerate*}}
\section{Adaptive Page Migration}
\label{s:promotion_demotion_demotion}

\begin{figure}[ht]
    \includegraphics{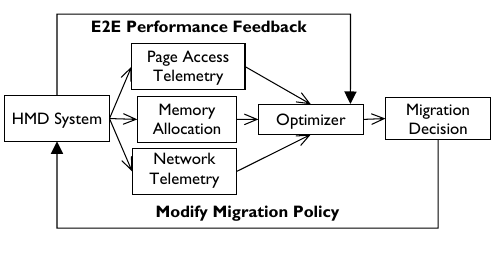}
\caption{Feedback loop to solve the optimization problem for page migration.}
\label{fig:feedback_loop}
\end{figure}

\BLUE{In Insight 3 in Section~\ref{ss:core_insights}, we showed that the optimal page migration policy is difficult to solve online due to the high computational complexity of O($\text{pages}^3$).}
We present two alternative approaches that can be used, depending on training data availability:
\begin{enumerate*}[label=(\alph*)]
    \item a network adaptive migration policy (as described in Section~\ref{ss:adaptive_page_promotion}), and
    \item a contextual multi-armed bandit agent that learns optimal migration policies trained with application performance metrics via a feedback loop guided by network telemetry, page telemetry, and memory allocations (as described in Sections~\ref{ss:reinforcement_learning} and \ref{ss:feedback_loop_optimization}).
\end{enumerate*}

\subsection{Network Adaptive Page Promotion}
\label{ss:adaptive_page_promotion}

We present a network-adaptive page promotion policy that does not require training data, based on Equation~\cref{eq:axiom_page_migration}.
By rearranging terms in the equation, we determine that pages should be promoted if \BLUE{${(\text{access}_p-\text{access}_d}) \times \text{Available Bandwidth} \times {\Delta \text{Latency} > \text{threshold}}$}.
The terms $\text{access}_p$ and $\text{access}_d$ (i.e., the access rates of the promoted and demoted pages) can be obtained at decision time using Page Telemetry, presented in Section~\ref{ss:access_access rate}.
The terms \text{Available Bandwidth} and $\Delta \: \text{Latency}$ can also be obtained online using network hardware performance counters with low overhead.

While this policy is simple to implement within an existing framework such as TPP, a few limitations prevent it from achieving optimal performance.
The policy does not take into account the dynamic nature of the application access patterns and adapt appropriately; the estimate of $\text{access}_d$ is stale, as demotion is not on the critical path of promotion; and
threshold selection is empirically guided (similar to other mechanisms, like Nimble and HeMem). 
To resolve these problems, we present the feedback loop approach with a contextual multi-armed bandit agent in Section~\ref{ss:reinforcement_learning}.

\subsection{Feedback Loop Formulation}
\label{ss:reinforcement_learning}

Figure~\ref{fig:feedback_loop} shows a feedback loop that can be used to solve the decision problem involved in page migration.
Again relying on Equation~\ref{eq:axiom_page_migration}, we derive features necessary for input to the feedback loop.
\begin{enumerate*}[label=(\alph*)]
    \item \emph{Page Access Telemetry:} $\text{access}_p$ (the access rate of the promoted page) is estimated using page access telemetry.
    \item \emph{Memory Allocation: } $\text{access}_d$ is dependent on the memory allocation. Larger values of memory allocation will lead to smaller $\text{access}_d$ values and vice versa.
    \item \emph{Network Telemetry: } The network cost term ($\frac{\text{page size}}{\text{bandwidth} \times \Delta \: \text{latency}}$) can be estimated from network telemetry (e.g., via network hardware performance counters). 
\end{enumerate*}
The migration decision computed by an optimizer simply decides whether the page should transfer from remote to local memory.
Application performance is used as feedback to improve the optimizer over multiple training runs.

However, that feedback loop is still impractical to use, as the page migration decision is made on a \emph{per-page} basis.
The optimizer is therefore difficult to scale, as every page migration requires an inference on the optimizer.
To resolve the problem, we changed the ``Migration Decision'' to a ``Migration Threshold,'' which is common to all pages.
Any page whose access rate is above the ``Migration Threshold'' is eligible for page promotion.
The feedback loop is now scalable, as it is independent of the number of pages used by the application.

The choice of optimizer can be varied, and there are multiple possible options.
We specifically chose contextual multi-armed bandits because  
\begin{enumerate*}[label=(\alph*)]
    \item it requires less hyperparameter tuning than reinforcement learning algorithms, such as PPO~\cite{ppo} and DDPG~\cite{ddpg}; and
    \item the inference overhead is smaller than that of Bayesian optimization.
\end{enumerate*}

\subsection{Feedback Loop Optimization}
\label{ss:feedback_loop_optimization}

\noindent \textbf{Solving the Feedback Loop.}
We model the solution to the feedback loop using a contextual multi-armed bandits (cMAB) agent.
cMAB~\cite{lu2010contextual} are a class of online decision-making problems in which an agent must choose from among several actions or ``arms'' to optimize a \textit{reward} given the current state of the environment defined as the \textit{context}.
Each episode is a single step corresponding to an application run.
We use Equation~\cref{eq:axiom_page_migration} to define the context, arms, and reward of the cMAB agent on a local node.

\noindent \textbf{Context. }
The context to the cMAB agent is equivalent to the input to the feedback loop in Figure~\ref{fig:feedback_loop} that includes page telemetry, memory allocation, and network telemetry.
By using migration thresholds (as discussed in Section~\ref{ss:reinforcement_learning}), we shift page telemetry to the actions and not use it in the context. 
Therefore, our context consists of features relevant to memory allocation and network telemetry:
\begin{enumerate*}[label=(\alph*)]
    \item \emph{Local Memory Allocation: } The total memory limit on the local node, which is limited by a user-defined \texttt{sysctl} variable.
    
    \item \emph{Local Node Memory Usage: } The maximum total memory used at the local node across the application run.

    \item \emph{Remote Memory Pool Usage: } The maximum total memory used at the remote node.
    
    \item \emph{Memory Network Traffic: } The total traffic sent on the memory network between the local node and remote memory pool.
\end{enumerate*}

\noindent \textbf{Actions. } The actions correspond to migration threshold selection that include:
\begin{enumerate*}[label=(\alph*)]
    \item \emph{Burst Duration Threshold} and
    \item \emph{Access Rate Threshold}.
\end{enumerate*}
Note that, an accessed page will be considered as a candidate for promotion only if the measured burst duration and access rate is above the corresponding threshold values.

\noindent \textbf{Reward.} We define the reward function for the cMAB agent using application performance as the primary optimization goal.
\BLUE{Improving application performance is equivalent to maximizing the access locality benefits described in Equation~\cref{eq:axiom_page_migration}.}
For batch jobs, we define application performance simply as the completion time.
On the other hand, for request-serving applications that do not finish, we mock up their completion times using the inverse of requests served per second. 
The final reward is the negative of the application completion time, to penalize higher completion times.

\begin{table}[]
\centering
\begin{tabular}{@{}ll@{}}
\toprule
\textbf{Hyperparameters} & \textbf{Value} \\ \midrule
num\_hidden\_layer        & 2              \\
hidden\_layer\_size        & 64             \\
learning\_rate           & 0.0005         \\
exploration\_fraction    & 0.1            \\
batch\_size              & 32             \\
MAX\_TRAIN              & 100000             \\\bottomrule
\end{tabular}
\caption{cMAB agent training hyperparameters.}
\label{tab:nn_agent}
\end{table}

\noindent \textbf{cMAB Implementation and Training.}
We implemented the cMAB agent based on a Deep Q-learning Network (DQN)~\cite{van2016deep} algorithm with $\gamma = 0$ to mimic a contextual multi-armed bandits solver.
The action space is discretized to use the thresholds as actions for the DQN solver.
Table~\ref{tab:nn_agent} shows the hyperparameters of the DQN agent used in our implementation and experiments with \sysname.
We selected the same hyperparameters as~\cite{mnih2015human} for faster policy training convergence.
We trained one cMAB agent with varying local and remote memory allocation configurations that ranged between 10\% and 90\% of the application's memory working set size.
The application was first run multiple times with maximal local memory allocation, and the cMAB agent was trained for \texttt{MAX\_TRAIN} episodes.
Next, the local memory allocation was gradually decreased with a step size of 10\% of the application working set size.
Each allocation was allowed to be trained for~\texttt{MAX\_TRAIN} episodes during the training of the cMAB agent.

\subsection{\sysname Kernel Implementation}
\label{ss:page_promotion}
\sysname promotes pages when NUMA hinting page faults arise (as described in Section~\ref{s:telemetry}).
Upon receiving a page fault, we update the burst duration and get the access rate estimate using the clustering algorithm described in Section~\ref{s:telemetry}.
\BLUE{The burst duration and access timestamps are recorded in the Linux kernel as an additional variable in \texttt{struct page} that uses 8 bytes.}
The overhead is negligible, as it corresponds to $<$0.1\% increase in total memory usage.
The burst duration and access rate are then compared against the threshold set by the cMAB agent as described in Section~\ref{ss:reinforcement_learning}.
In addition, we also implemented a \texttt{LOW\_WATERMARK} variable that prevents further pages from being migrated if no space is available on the local node.
The \texttt{LOW\_WATERMARK} variable denotes a hard threshold to prevent the local node memory from exceeding its allocation. 
If a page satisfies  the \texttt{LOW\_WATERMARK} and meets the cMAB-set thresholds, an attempt is made to transfer the page from the remote node to the local node by reusing the existing \texttt{migrate\_pages()} function provided by the Linux kernel tools for NUMA systems.

We reused the existing Linux disk paging implementation for page demotion using the \texttt{kswapd} daemon.
Instead of swapping to the disk, pages are demoted to the remote node.
We wake up the \texttt{kswapd} daemon if memory allocation on the local node exceeds a \texttt{HIGH\_WATERMARK}.
\texttt{HIGH\_WATERMARK} is set approximately 10 MB higher than \texttt{LOW\_WATERMARK} to allow for slack to enable page promotion if the local node's memory usage is reaching its allocation capacity.
\section{Evaluation}
\label{s:results}

Our evaluation aims to address the following aspects:
\begin{enumerate}[label=(\alph*),leftmargin=22pt]
\item Variance in performance gains from \sysname across applications,
\item \sysname's performance in network contention scenarios,
\item \sysname's performance when pretrained on a similar application,
\item \sysname's training time and performance degradation during training,
\item Breakdown of performance gains by each \sysname component, and
\item The overhead cost of running \sysname along with the HMD system.
\end{enumerate}

\subsection{System Setup}
\label{ss:system_setup}
We used ThymesisFlow~\cite{thymesisflow}, an open-source prototype for hardware memory disaggregation. 
ThymesisFlow implements a hardware-software co-designed memory disaggregation interconnect on top of the POWER9 architecture by directly interfacing the memory bus via the OpenCAPI cache-coherent interconnect protocol.
Each node featured a dual-socket POWER9 CPU (32 physical cores and 128 parallel hardware threads) and 512 GB of RAM.
Each node is equipped with an AlphaData 9V3 card that featured a Xilinx Ultrascale FPGA that implemented the OpenCAPI stack and the ThymesisFlow logic.
Our prototype implements the HMD system described in Section~\ref{ss:background_hmd}, in which the memory network has a shared 100 Gb/s link to the remote memory pool with \BLUE{$\sim$900ns access latency}~\cite{pinto2020thymesisflow}. 
Network contention is observed on this shared link.

\subsection{Experiment and Workload Setup}
\label{ss:workload}

To evaluate \sysname, we selected the memory-intensive applications shown in Table~\ref{tab:workloads}. 
These applications are representative of common workloads in cloud datacenters or HPC systems.

\begin{table}[]
\resizebox{\linewidth}{!}{
\begin{tabular}{@{}lll@{}}
\toprule
Application  & Memory Usage & Description                                                                                                                                                                                         \\ \midrule
Redis~\cite{carlson2013redis}        & 16 GB        & \begin{tabular}[c]{@{}l@{}}In-memory data structure store,\\ run with memtier benchmark with\\  10,000 requests/s  \end{tabular} \\ \\
Graph500~\cite{murphy2010introducing}     & 48 GB        & \begin{tabular}[c]{@{}l@{}}Multiple iterations of Breadth First \\ Search (BFS)\\ and Single Source Shortest Path (SSSP), \\ configuration uses problem size = 25,\\  edge factor = 16\end{tabular} \\ \\
voltDB~\cite{stonebraker2013voltdb}       & 12 GB        & \begin{tabular}[c]{@{}l@{}}In-memory and fully ACID-compliant\\  RDBMS with a share-nothing architecture, \\ run with TPC-C benchmark\end{tabular}        \\   \\
HashJoin~\cite{zeller1990adaptive}     & 36 GB        & \begin{tabular}[c]{@{}l@{}}Database join algorithm that uses\\  hashcodes in equijoins\end{tabular}                                                                                                 \\  \\
ML Inference~\cite{vaswani2017attention} & 100 GB       & \begin{tabular}[c]{@{}l@{}}Transformer-based model inference\end{tabular}  \\                                                                                                          \\
AMG~\cite{stuben1983algebraic}          & 40 GB        & \begin{tabular}[c]{@{}l@{}}Parallel algebraic multigrid solver that \\ represents common operations in HPC systems\end{tabular}                                                                     \\ \bottomrule
\end{tabular}
}
\caption{Applications and their configurations used in evaluating \sysname.}
\label{tab:workloads}
\end{table}


\noindent \textbf{Experiment Methodology.}
We ran experiments with five page migration mechanisms: \sysname, \sysname Network Adaptive Policy (which does not require training), MEMTIS, TPP, and Nimble.
The applications were run with different local-to-remote memory allocation ratios varying between 10--90\%.
In addition to these page migration mechanisms, we also ran the applications in the ``None'' configuration which denotes only remote memory was used.
Contention on the memory disaggregation fabric was modeled by executing the STREAM benchmark concurrently with applications.
\sysname was trained on each application separately, using the contextual multi-armed bandit agent described in Section~\ref{ss:reinforcement_learning}.
To train each application, we decremented the local memory allocation by 10\% after every \texttt{MAX\_TRAIN} episodes. 
The agent learned and adjusted its actions (the cutoff for burst duration and access rate) for each new allocation.
To avoid unnecessary application execution during training, we also cached the context-to-reward evaluation.
Specifically, we cache the mapping from the context, access rate, and burst duration cutoffs to application runtime after application execution.
When the same context and action set was encountered again, we could simply retrieve it from the saved mapping instead of rerunning the application.
Without the caching mechanism, the cMAB training algorithm would run each application for \texttt{MAX\_TRAIN} episodes.
However, the introduction of caching reduced the number of application runs to $\sim$1000, resulting in a $\sim$100\texttimes~reduction in training time.


\begin{figure}[h]
    \centering
    \includegraphics{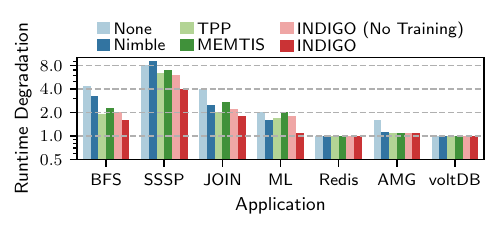}
    \caption{Per-application runtime degradation across page migration algorithms for 10\% local memory allocation.}
    \label{fig:app_runtime_degradation_low}
\end{figure}%

\begin{figure}[h]
    \centering
    \includegraphics{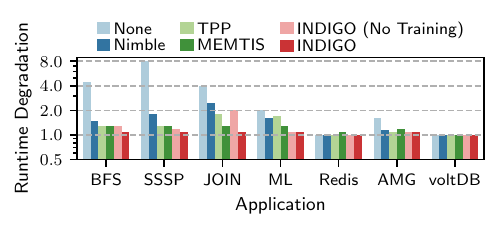}
    \caption{Per-application runtime degradation across page migration algorithms for 80\% local memory allocation.}
    \label{fig:app_runtime_degradation_high}
\end{figure}%
\subsection{Evaluation Results}
\label{ss:eval-results}

\begin{figure}
  \centering
  \begin{tikzpicture}
    \node[anchor=south west,inner sep=0] (image) at (0,0) {\includegraphics{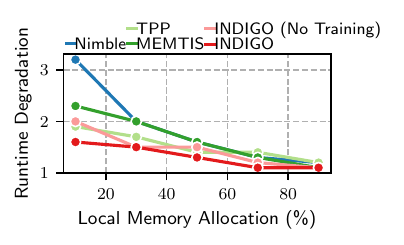}};
    \fill [white] (1,3.55) rectangle (1.95,4);
  \end{tikzpicture}
  \caption{BFS runtime degradation with increasing local memory allocations for different page migration mechanisms.}
  \label{fig:allocation_variance}
\end{figure}

\noindent \textbf{Application Performance.}
We find that the performance benefit from page migration depends greatly on the application characteristics and the local-to-remote memory allocation ratio.
Figures~\ref{fig:app_runtime_degradation_low} and \ref{fig:app_runtime_degradation_high} show the variation in applications' runtime degradation when they were running with \sysname, Nimble, TPP, MEMTIS, and page migration completely disabled.
We define runtime degradation as the ratio between the observed runtime and the runtime with 100\% local memory allocation (i.e., the lowest possible runtime for the application).
Each application was run in two configurations:
\begin{enumerate*}[label=(\alph*)]
\item with low local memory (10\% of the total application working set size), and 
\item with high local memory (80\% of the total size).
\end{enumerate*}

In addition, our evaluation confirms several hypotheses about the relationship among page migration performance, memory allocations, and application characteristics.
\begin{itemize}[wide=0pt]
    \item Overall, we find that \sysname offers a runtime improvement of up to 50--70\% compared to other page migration mechanisms. Runtime with \sysname (No Training) is comparable to other mechanisms without network contention, but offers up to 30--50\% improvement under network contention (as we describe later in the evaluation).   
    \item Higher local memory allocations result in less runtime degradation. For example, when comparing Figure~\ref{fig:app_runtime_degradation_low} and Figure~\ref{fig:app_runtime_degradation_high} across applications, we find that increasing the local memory allocation reduces runtime degradation.
    The lower runtime can be explained by the smaller average memory access latency as additional pages are placed in local memory.
    \item Applications with high computational overhead, such as Redis\footnote{Redis communicates via TCP sockets which becomes the dominant cost, and remote memory access latency is effectively hidden.} and voltDB, do not experience significant performance degradation from the use of remote memory. In consequence, none of the page migration mechanisms are able to provide a significant benefit, as these applications are mostly computing on CPU rather than accessing memory.
    \item While page migration with \sysname can reduce runtime degradation significantly, it may not be sufficient to close the gap between the baseline (full local memory allocation) and partial local memory allocation. This is especially true for applications with more randomized access patterns, such as graph traversal (BFS and SSSP). However, given the constraint of limited local memory, \sysname still outperforms other mechanisms.
    \item We find that for numerous applications and local memory allocation combinations, application performance can reach nearly the baseline.
    Such improvement is observed when the application performs significant computation on the migrated pages (for e.g. layer-by-layer inference in case of ML), hence appropriate page migration is sufficient to completely hide remote access latency. 
    \item Figure~\ref{fig:allocation_variance} shows that \sysname's advantage over other page migration mechanisms, such as Nimble, TPP, and MEMTIS, is more pronounced for smaller local memory allocations.
    This observation can be explained by the left side of Equation~\cref{eq:axiom_page_migration}.
    As the local memory allocation decreases, the access rate of the demoted page ($\text{access}_d$) increases, and the corresponding optimality threshold of $\text{access}_p$ also increases.
    Therefore, fewer pages benefit from promotion, and \sysname's selectivity in page promotion becomes critical.
    Specifically, we find that \sysname promotes between 20--40\% less pages compared to other page migration mechanisms.
\end{itemize}

\begin{figure}[ht]
    \includegraphics{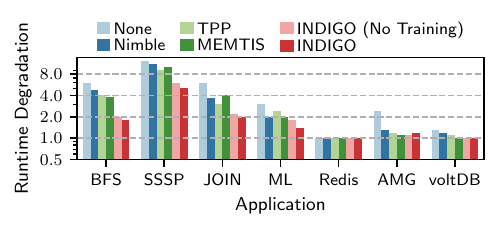}
    \caption{Per-application runtime degradation during network contention for 10\% local memory allocation.}
    \label{fig:app_runtime_network_contention}
\end{figure}

\begin{figure}[ht]
    \includegraphics{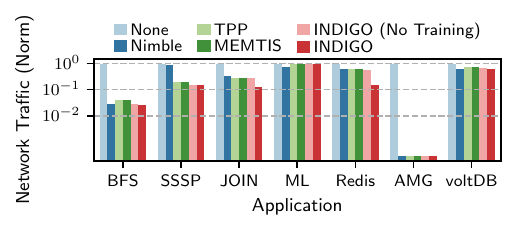}
    \caption{Per-application network traffic across page migration algorithms for 10\% local memory allocation.}
    \label{fig:app_interconnect_utilization_low}
\end{figure}

\noindent \textbf{Impact of Network Contention.}
To understand the impact of network contention on the migration mechanisms, we used the STREAM benchmark (that performs serial memory copy operations via the COPY kernel) to produce memory traffic.
STREAM combined with other applications generates network contention by utilizing 50\% of the available link bandwidth.
Such network contention corresponds to bandwidth contention observed in commercial HMD systems~\cite{h3-platform}.
Figure~\ref{fig:app_runtime_network_contention} shows the impact on runtime degradation for each application and page migration mechanism under such network contention.
We found that both versions of \sysname performed better than other mechanisms.
In particular, we observed that the performance of \sysname (No Training) was significantly better than that of TPP, Nimble, and MEMTIS.
For instance, the performance of the BFS application \sysname (No Training) was comparable to that of TPP without network contention but provided a 50\% performance improvement under network contention.
The improved performance of both versions of \sysname can be explained by the right side of Equation~\cref{eq:axiom_page_migration}.
\sysname considers the network cost term ($\frac{\text{page size}}{\text{bandwidth} \times \Delta \: \text{latency}}$) and lowers the promotion rate by setting a higher cutoff for the $\text{access}_p$ estimate, thereby satisfying the optimal migration condition.
For example, when running BFS, \sysname transfers 12000 pages/sec under contention, compared to TPP that transfers 22000 pages/sec.
Promotion rates for other mechanisms are shown in Figure~\ref{fig:insight_promotion_rate}.

\noindent \textbf{Network Traffic Reduction.}
We found that page migration using \sysname results in less total network traffic across applications than other page migration mechanisms.
Figure~\ref{fig:app_interconnect_utilization_low} shows the normalized network traffic (cache miss and page migration traffic) for different applications and page migration algorithms with local memory allocation equal to 10\% of the total application working set size.
Traffic is normalized with respect to the full remote memory allocation that consumes the most network bandwidth.
Total network traffic using \sysname is up to 4000\texttimes~lower (in case of AMG) and 2.3\texttimes~lower (in case of JOIN) than can be achieved using only remote memory and other page migration algorithms.
Again, the reason is the cumulative effect of traffic from page promotion and cache miss.
Although the page promotion traffic is higher, the total number of cache misses decreases as fewer pages are placed in local memory.
Hence, the overall network traffic (a combination of migration and access traffic) is lower with \sysname.
In addition, we found that the applications with the highest total network traffic may not have the worst runtime degradation, as the memory access cost can be masked by high computational costs (for example, due to expensive matrix multiplications in  ML inference workloads or high network processing costs in the case of Redis).

\begin{figure}
\hfill
\begin{minipage}{0.23\textwidth}
\centering
\includegraphics[width=1.6in]{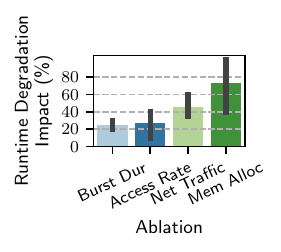}
\caption{Ablation study of \sysname components.}
\label{fig:ablation_study}
\end{minipage}
\hfill
\begin{minipage}{0.23\textwidth}
\centering
\includegraphics[width=1.6in]{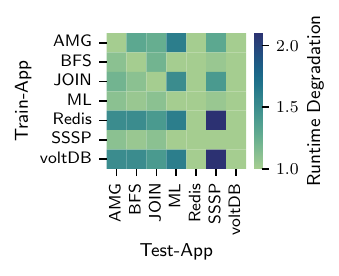}
\caption{\sysname performance transfer across applications without retraining.}
\label{fig:transfer}
\end{minipage}
\end{figure}

\noindent \textbf{Ablation Study of \sysname Components.}
We performed an ablation study to understand the contribution of each \sysname component on end-to-end performance improvement.
To do so, we modified \sysname by disabling each component one at a time and comparing the drop in performance relative to the original mechanism.
The four components we identified for the ablation study were
 \emph{burst duration}, which accounts for dynamic application access patterns and decision context ($\Delta T$);
 \emph{access rate}, which accounts for the access rate estimate ($\text{access}_p$ and $\text{access}_d$);
  \emph{network traffic}, which accounts for the network cost term ($\frac{\text{page size}}{\text{bandwidth} \times \Delta \: \text{latency}}$); and
  \emph{memory allocation}, which, along with the access rate of the promoted page ($\text{access}_p$), can improve the estimate of the access rate of the demoted page ($\text{access}_d$).
Figure~\ref{fig:ablation_study} shows the reduction in gains due to the use of INDIGO, when each component is disabled.
We found that all \sysname components contribute significantly to overall performance gains, with memory allocation being the most significant.

\begin{table}
\centering
\begin{tabular}{ccccc}
\toprule
Applications & INDIGO & Nimble & TPP & MEMTIS\\
\midrule
SSSP & \textbf{6.2}\texttimes & 9\texttimes & 7.3\texttimes & 7\texttimes\\
JOIN & \textbf{2.5}\texttimes & 3.4\texttimes & 3.2\texttimes & 2.8\texttimes\\
voltDB & 1\texttimes & 1\texttimes & 1\texttimes & 1\texttimes \\
\bottomrule
\end{tabular}
\caption{Application runtime degradation across page migration algorithms in a multi-tenant setting.}
\label{tab:multitenancy}
\end{table}

\noindent \textbf{Multi-tenant Performance.}
\sysname can be extended to work in the multi-tenant setting where multiple applications on the local node are simultaneously migrating pages from the remote memory pool.
There are two aspects to the \sysname's extension in the multi-tenant setting: 
\begin{enumerate*}[label=(\alph*)]
\item the underlying network cost model from Equation~\cref{eq:axiom_page_migration}, and 
\item the Linux kernel implementation.
\end{enumerate*}
The cost model continues to hold for each application's page migration performance with modifications in the $\text{bandwidth}$ and $\Delta \text{latency}$ terms to account for contention between applications.
Therefore, we can use the heuristic policy and trained cMAB model with per-application memory traffic measurements.
The Linux kernel also needs to be modified to support per-application memory allocation, memory usage, and thresholds for access rate and burst duration.
We support these per-application metrics by creating an additional task-metric mapping using the task struct within the kernel space.

To evaluate \sysname in the multi-tenant setting, 
we choose the following three applications: (a) SSSP, (b) JOIN, and (c) voltDB.
We choose SSSP as it is the most performance sensitive to page migration.
voltDB and JOIN were chosen as they were the only other applications that prevent CPU core oversubscription.
We avoid CPU core overscubscription, as in the event of oversubscription, page migration would not be the major cause of performance degradation.
All applications were run with 10\% local memory allocation.
Table~\ref{tab:multitenancy} shows the runtime degradation across page migration algorithms compared to an isolated full local memory allocation case.
Overall, we find that the performance benefits of \sysname continue to hold.

\noindent \textbf{cMAB Transfer across Applications.}
We found that \sysname performance transferred without retraining across similar applications.
Figure~\ref{fig:transfer} shows the runtime degradation when \sysname was trained on the \emph{Train-App} and tested against the \emph{Test-App}.
We found that for similar applications, such as BFS and SSSP, the performance of a trained cMAB agent transferred with less than 5\% degradation. 
On the other hand, applications that have low variability in performance (such as voltDB and Redis) are bad candidates for training, as they do not allow any useful learning.
In addition, we note that use of a pre-trained application model limited the maximum amount of degradation to only twice that of the baseline degradation.
Thus, a pre-trained model can be used to kickstart the re-training and avoid high initial runtime degradation.

\noindent \textbf{Runtime Overhead.}
\sysname introduces low runtime and memory overhead compared to complete disabling of page migration.
Runtime overhead is primarily introduced from the extra computational cost associated with page marking and page faults and in the worst-case is limited to 10\% of the application runtime in our experiments.
The runtime overhead is not specific to \sysname, as other page migration mechanisms, such as TPP and Nimble, use identical page poisoning systems.
Memory overhead is introduced by modifying the page struct data structure to record the burst duration. 
The memory overhead corresponds to an increase of 8 bytes for every page in the operating system (i.e., 8 bytes for 64 KB $\sim$0.01\%).
\section{Discussion}
\label{s:discussion}


\textbf{Difference between CXL and OpenCAPI.}
\sysname uses ThymesisFlow~\cite{thymesisflow}, which instantiates an HMD system using the OpenCAPI coherent interconnect protocol.
However, we expect that most future research activity targeting system composability or component disaggregation will focus on CXL because the OpenCAPI consortium has donated all of its IPs to the CXL Consortium~\cite{ocapi_cxl}.
OpenCAPI~\cite{openCAPI} and CXL~\cite{CXL} have some fundamental physical differences but share most of the concepts related to memory access and operating system support.
Both OpenCAPI and CXL enable manipulation of memory external to a machine by means of a dedicated CPU-less NUMA node, alongside local NUMA nodes hosting one or more processors and their local memory.
Also, in both cases, remote memory is accessed at the cache line granularity.
Therefore, \sysname and any page migration mechanism that moves pages across NUMA nodes are applicable to both technologies.

\noindent \textbf{\sysname Scalability to Multiple Remote Memory Pools.}
\sysname considers all remote memory as a single NUMA node and relies on the hardware implementation to implement multiple remote memory pools.
In addition, the degradation in access latency when moving from local to remote memory is 10--100\texttimes~\cite{patke2022evaluating,gouk2022direct,pinto2020thymesisflow}, which accounts for a much larger slowdown than that caused by variation within remote memory pools.
Hence, we expect that the performance of \sysname will not be significantly affected by the choice of remote memory pool that will be the candidate for page promotion and demotion.



\section{Related Work}
\label{s:related_work}

\noindent \textbf{Page Migration.}
Prior work has explored hardware-assisted~\cite{meswani2015heterogeneous,mogul2009operating,ramos2011page,lihopp} and application-guided~\cite{dulloor2016data,wei2015exploiting,giardino2016soft2lm} page migration for tiered memory systems, which may not be practical as they require specialized hardware support or application redesign from the ground up.
A large body of work exists in page migration for disaggregated and tiered memory that is independent of the application and hardware support~\cite{yan2019nimble,liu2019hierarchical,kim2021exploring,maruf2022tpp,agarwal2017thermostat,maruf2022multi,ren2023hm,moon2023adt,giannoula2023daemon,kim2021rapidswap,doudali2021cori}.
We discuss the limitations of these approaches in Section~\ref{ss:page_migration_characterization}.

\noindent \textbf{Memory Disaggregation.}
Memory disaggregation exposes capacity available in remote hosts as a pool of memory shared among many machines.
OS-based~\cite{shan2018legoos,gu2017efficient,amaro2020can,al2020effectively} or runtime-based memory disaggregation~\cite{aguilera2018remote,ruan2020aifm,wang2020semeru}
is designed for RDMA over InfiniBand or Ethernet networks.
However, the access latency is an order of magnitude higher than for hardware-based approaches.
Recently, with the introduction of interconnect protocols, there has been a rise in HMD systems that directly connect the processor and remote memory via the cache-line such as Pond~\cite{li2023pond}, DirectCXL~\cite{gouk2022direct}, and ThymesisFlow~\cite{pinto2020thymesisflow}.
\sysname can be used in conjunction with such HMD systems to mitigate the impact of remote memory access latency.

\section{Conclusion}
\label{s:conclusion}

We present \sysname, a network-aware page migration framework for HMD systems.
At the core of \sysname is an adaptive page promoter that uses novel page telemetry metrics, such as access rate and burst duration, to select pages dynamically for promotion from remote to local memory via a contextual multi-armed bandit agent.
Evaluation of \sysname on a real HMD system with common cloud and HPC workloads shows up to 50--70\% improvement in application performance.
\sysname and further research on page migration is crucial for the reduction of the TCO of infrastructure, by enabling data center providers to deploy cost-effective and high-capacity shared memory pools. 

\bibliographystyle{ACM-Reference-Format}
\bibliography{bibliography}

\end{document}